\documentclass[secnumarabic, twocolumn, graphics,floatfix, superscriptaddress, tightenlines, aps, prb]{revtex4-2}
\usepackage[dvips]{graphicx}
\usepackage{amssymb,amsmath}
\usepackage{nccmath}
\DeclareMathOperator{\Tr}{Tr}
\usepackage{dcolumn}
\usepackage{bm}
\usepackage{color}
\usepackage{scrextend}
\usepackage{hyperref}
\usepackage[sort&compress]{natbib}
\usepackage{multirow}
\usepackage{hhline}

\begin{document}

\title{Deep optical cooling of coupled nuclear spin-spin and quadrupole reservoirs in a GaAs/(Al,Ga)As quantum well}

\author{M.~Kotur}
\affiliation{Experimentelle Physik 2, Technische Universität Dortmund, D-44221 Dortmund, Germany}

\author{D.~O.~Tolmachev}
\affiliation{Experimentelle Physik 2, Technische Universität Dortmund, D-44221 Dortmund, Germany}

\author{V.~M.~Litvyak}
\affiliation{Spin Optics Laboratory, St.~Petersburg State University, 198504 St.~Petersburg, Russia}

\author{K.~V.~Kavokin}
\affiliation{Spin Optics Laboratory, St.~Petersburg State University, 198504 St.~Petersburg, Russia}

\author{D.~Suter}
\affiliation{Experimentelle Physik 3, Technische Universität Dortmund, D-44221 Dortmund, Germany}

\author{D.~R.~Yakovlev}
\affiliation{Experimentelle Physik 2, Technische Universität Dortmund, D-44221 Dortmund, Germany}
\affiliation{Ioffe Institute, Russian Academy of Sciences, 194021 St.~Petersburg, Russia}

\author{M.~Bayer}
\affiliation{Experimentelle Physik 2, Technische Universität Dortmund, D-44221 Dortmund, Germany}
\affiliation{Ioffe Institute, Russian Academy of Sciences, 194021 St.~Petersburg, Russia}

\begin{abstract}
The selective cooling of \textsuperscript{75}As spins by optical pumping followed by adiabatic demagnetization in the rotating frame is realized in a nominally undoped GaAs/(Al,Ga)As quantum well. The rotation of 6 kG strong Overhauser field at the \textsuperscript{75}As Larmor frequency of 5.5 MHz is evidenced by the dynamic Hanle effect. Despite the presence of the quadrupole induced nuclear spin splitting, it is shown that the rotating \textsuperscript{75}As magnetization is uniquely determined by the spin temperature of coupled spin-spin and quadrupole reservoirs. The dependence of heat capacity of these reservoirs on the external magnetic field direction with respect to crystal and structure axes is investigated. The lowest nuclear spin temperature achieved is 0.54 $\mu$K, which is the record low value for semiconductors and semiconductor nanostructures.

\end{abstract}

 \maketitle

\section{Introduction}
\label{sec:intro}
The hyperfine interaction is one of the main sources of spin relaxation and decoherence of charge carriers in semiconductor structures, which hinder their application in spintronics. This problem is especially strong in the most technologically versatile family of heterostructures based on compounds of the III and V groups of the periodic table, where all the nuclear species have nonzero spins. One of the possible approaches to solve this problem is to develop efficient methods of cooling the nuclear spin system down to the temperature of phase transition into the antiferromagnetically ordered state \cite{goldman1974principles,oja1997nuclear}. The general principle of cooling the nuclear spin system is to adiabatically demagnetize initially polarized nuclear spins by lowering the external magnetic field. The higher the initial polarization, the lower spin temperature of nuclei can be reached \cite{goldman1970spin}.

The possibility to dynamically polarize nuclear spins in a semiconductor via their hyperfine coupling with optically oriented electron spins opens ways to realization of deep nuclear spin cooling by minimal technical means, i.e. to avoid using dilution refrigerators and/or high magnetic fields. The key point here is the choice of the optimal structure. The minimal list of necessary prerequisites includes the possibility to optically polarize nuclear spins to a high degree and long nuclear spin-lattice relaxation. These conditions are realized in nominally undoped GaAs/(Al,Ga)As quantum wells \cite{mocek2017high}, making these structures prospective for further investigation of the properties of their nuclear spin system under deep cooling.

The possibility to move down along the spin temperature scale depends on the interactions in which nuclear spins are involved. These include, in addition to the Zeeman interaction with the external field, the dipole-dipole interaction between magnetic moments of nuclei, and their indirect coupling via electron states \cite{abragam1961principles}. All the interactions except the Zeeman one are usually lumped together under the name of spin-spin interactions, which form the spin-spin energy reservoir. In addition, if nuclei have spins larger than $\frac{1}{2}$, as it is the case in III-V semiconductors, they experience quadrupole coupling with electric field gradients induced by strain \cite{abragam1961principles,eickhoff2003mapping,flisinski2010optically}. In case of strong quadrupole splitting (e.g. in self-assembled quantum dots) it may prevent establishing of the thermodynamic equilibrium in the nuclear spin system \cite{maletinsky2009breakdown}. However, if the quadrupole and spin-spin interaction energies per nucleus are comparable, a quadrupole and spin-spin energy reservoirs are effectively coupled, and the nuclear spin system can be characterized by a unified spin temperature \cite{vladimirova2018spin}. The nuclear magnetic ordering is expected to develop when the coupled energy reservoirs are cooled down below a certain critical spin temperature. For this reason, understanding the properties of the spin-spin and quadrupole (SS\&Q) reservoir under cooling is crucial for realization of nuclear magnetic ordering in a specific structure.

The SS\&Q reservoir can be cooled either together with the Zeeman reservoir, or separately. The latter option is realized by adiabatic demagnetization in the rotating frame (ADRF) \cite{slichter1961adiabatic,anderson1962nuclear,redfield1969nuclear}.  Within this approach, the static external magnetic field is kept unchanged, while the nuclear spins are manipulated by variation of amplitude and/or frequency of the applied radiofrequency field. The ADRF method allows one to considerably reduce the spin-lattice relaxation rate and to address independently the spins of specific isotopes in a multi-isotope crystal like GaAs.

In this work, we use ADRF with optical pumping of nuclear spins and optical detection of the free induction decay (FID) signals to selectively cool the \textsuperscript{75}As spins in a GaAs/(Al,Ga)As quantum well. We demonstrate that the \textsuperscript{75}As spin polarization in the rotating frame is well described by the spin temperature theory, and estimate the contributions of dipole-dipole and quadrupole interactions into the heat capacity of the SS\&Q reservoir. Cooling of the \textsuperscript{75}As SS\&Q reservoir down to $\approx0.55$ $\mu$K is realized for both positive and negative spin temperatures.

\section{Theoretical background}
\label{sec:theory}

The adiabatic demagnetization in a rotating frame can be realized in several ways, see Ref.~\cite{goldman1970spin}. Here we use the spin lock technique, known to be one of the methods ensuring lowest entropy gain. At the first stage of the experiment, the nuclear spins are polarized along the external magnetic field. The higher the initial polarization, the lower spin temperature can be reached. In semiconductors, high nuclear polarization can be reached by optical pumping mediated by hyperfine interaction with photoexcited electrons~\cite{meier2012optical}. The high spin polarization parallel to the external field can be interpreted as a result of cooling of the nuclear spin system to a temeperature which is lower than that of the lattice by absolute value, and can be either positive or negative, depending on whether nuclear spins are polarized along the external field or opposite to it.  At the second stage, a $\frac{\pi}{2}$ radio-frequency pulse is applied at the Larmor frequency of the selected isotope (in our case \textsuperscript{75}As), which tips the mean spin vector of the isotope so that it becomes perpendicular to the external field. 
Once the mean spin is tipped at 90$^\circ$ to the external field, the populations for all the Zeeman sublevels become equal. Correspondingly,  the temperature of the Zeeman reservoir of \textsuperscript{75}As becomes infinite. However, the entropy of the spin system remains low as the coherence of the tipped spins is maintained. Within the spin lock protocol, this is done by switching on the radiofrequency (RF) field at the Larmor frequency, called locking field, which differs in phase by $\frac{\pi}{2}$ from that of the tipping pulse. In the coordinate frame rotating together with the tipped nuclear spin around the static external field, the locking field has a static component parallel (or antiparallel, depending on the sign of the phase shift of the locking field) to the mean nuclear spin. The decay of the rotating nuclear magnetic moment must, therefore, be accompanied by relaxation of its Zeeman energy in the locking field, which profoundly slows it down. Instead of the spin-spin relaxation time $T_2$ of about 0.1 millisecond, the decay of the rotating spin polarization in the locking field occurs on the scale of the spin-lattice relaxation time in the rotating frame $T_{1\rho}$ \cite{vanderhart197913c}, which can amount to many seconds. This is a manifestation of the quasi-equilibrium nature of this long-living spin polarization, which is characterized by the spin temperature of the SS\&Q reservoir $\Theta_N$. If the amplitude of the locking field, $b$, is changed slowly (that is, $\frac{db}{dt}<\frac{B_L}{T_2}$), starting from its initial value $b_0$, $\Theta_N$  should follow the adiabatic curve given by the well-known equation~\cite{goldman1970spin}:
\begin{equation}
\Theta_N(b)=\Theta_N(b_0)\sqrt{\frac{b^2+B_L^2}{b_0^2+B_L^2}},
\label{eq:temperature}
\end{equation}
where $B_L$ is the local magnetic field due to spin-spin and quadrupole interactions. Although the magnetic field rate constrain is only relevant when $b$ is comparable to $B_L$, during our experiment the magnetic field was changed so that the given condition was always fulfilled even though the initial value of $b_0$ exceeded $B_L$ several times. The mean spin of the tipped isotope is determined by $\Theta_N$ and $b$:
\begin{equation}
\langle I \rangle=\frac{I(I+1)\hbar\gamma_Nb}{3k_B\Theta_N},
\end{equation}
where $I$ and $\gamma_N$ are spin and gyromagnetic ratio of the tipped isotope and $k_B$ is the Boltzmann constant.

The Overhauser field acting upon the electron spins is proportional to the mean nuclear spin. In particular, the amplitude of the rotating field of the tipped isotope is:
\begin{equation}
\begin{split}
&B_\perp(b)=\frac{A_N\langle I \rangle}{\hbar\gamma_e}=\frac{A_NI(I+1)\gamma_Nb}{3k_B\Theta_N(b)\gamma_e}=\\
&=\frac{A_NI(I+1)\gamma_Nb}{3k_B\Theta_N(b_0)\gamma_e}\sqrt{\frac{b_0^2+B_L^2}{b^2+B_L^2}}=B_\perp(b_0)\frac{b}{b_0}\sqrt{\frac{b_0^2+B_L^2}{b^2+B_L^2}},
\end{split}
\label{eq:B_perp}
\end{equation}
where $A_N$ is the hyperfine coupling constant for the tipped isotope and $\gamma_e$ is the electron gyromagnetic ratio. The only parameter that determines the shape of the function $B_\perp(b)$ is the local field $B_L$, which is in fact the measure of heat capacity of the SS\&Q reservoir. It is defined by the relation:
\begin{equation}
B_L^2=\frac{\Tr[\hat{H}_{ss}^2+\hat{H}_Q^2]}{\hbar^2\gamma_N^2\Tr[\hat{I}_B^2]},
\label{eq:local}
\end{equation}
where $\hat{H}_{ss}$ and $\hat{H}_Q$ are the secular, i.e. commuting with the Zeeman Hamiltonian, parts of the spin-spin and quadrupole Hamiltonian correspondingly, and $\hat{I}_B$ is the operator of the projection of nuclear spin on the external field \cite{goldman1970spin}.

As distinct from the case of adiabatic demagnetization in the laboratory frame, $B_L$ depends on the orientation of the static external field with respect to the crystal and structure axes. The quadrupole splitting of the nuclear spin energy levels results from the electric field gradient (EFG), which interacts with the quadrupole moments of the nuclei. The EFG is zero in unperturbed cubic lattices. In GaAs, it arises due to strain and electric fields. In thin planar structures like ours, one can safely assume that shear strain in the XZ and YZ planes (where Z is the structure growth axis) is zero. The remaining components of the strain tensor $\varepsilon$ (that is, uniaxial along Z and biaxial in the XY plane), together with the electric field $E$ that could arise along Z due to spatial or surface charge, result in the following general form of the quadrupole Hamiltonian:
\begin{equation}
\begin{gathered}
\hat{H}_Q=\frac{E_{Qz}}{2}\left[\hat{I}_z^2-\frac{I(I+1)}{3}\right]+\\ +\frac{E_{QR}}{2\sqrt{3}}(\hat{I}_x^2-\hat{I}_y^2)+\frac{E_{QI}}{2\sqrt{3}}(\hat{I}_x\hat{I}_y+\hat{I}_y\hat{I}_x),
\end{gathered}
\label{eq:H_Q}
\end{equation}
where the energies $E_{Qz}$, $E_{QR}$ and $E_{QI}$ are related to the strain tensor components and the electric field by the material tensors $S$ and $R$:
\begin{equation}
\begin{gathered}
E_{Qz}=\frac{3eQS_{11}}{I(I+1)}\left(\varepsilon_{zz}-\frac{\varepsilon_{xx}+\varepsilon_{yy}+\varepsilon_{zz}}{3}\right) \\
E_{QR}=\frac{3\sqrt{3}eQS_{11}}{I(I+1)}\left(\varepsilon_{xx}-\varepsilon_{yy}\right) \\
E_{QI}=\frac{3\sqrt{3}eQS_{44}}{2I(I+1)}\varepsilon_{xy}+\frac{3\sqrt{3}eQR_{14}}{2I(I+1)}E,
\end{gathered}
\end{equation}
where $e$ is the electron charge and $Q$ is the quadrupole moment of the nucleus. The angle between the principal axis of the EFG tensor in the plane and the [100] crystal axis, $\zeta$, is defined by the relations:
\begin{equation}
\begin{gathered}
\cos2\zeta=\frac{E_{QR}}{\sqrt{E_{QR}^2+E_{QI}^2}} \\
\sin2\zeta=\frac{E_{QI}}{\sqrt{E_{QR}^2+E_{QI}^2}}.
\end{gathered}
\end{equation}
In a strong static magnetic field, $B\gg(\hbar\gamma_N)^{-1}\sqrt{E_{Qz}^2+E_{Q\perp}^2}$, with the direction defined by the polar angle $\theta$ and the azimuthal angle $\alpha$, the secular part of the quadrupole Hamiltonian in Eq. \eqref{eq:H_Q} can be written as: 
\begin{equation} \medmath{
\begin{split}
&\hat{H}_Q=\frac{1}{4}\left[E_{Qz}(3\cos^2\theta-1)+E_{Q\perp}\sqrt{3}\sin^2\theta \cos2(\zeta-\alpha)\right]\times \\
&\times\left[\hat{I}_B^2-\frac{I(I+1)}{3}\right],
\end{split}}
\end{equation}
where $\theta$ and $\alpha$ are the angles between the external magnetic field and structure and crystal axes of the sample, respectively and $E_{Q\perp}=\sqrt{E_{QR}^2+E_{QI}^2}$. The quadrupole contribution to the local field then reads:
\begin{equation} \medmath{
\begin{split}
B_{LQ}&=\left[\frac{\Tr[\hat{H}_Q^2]}{\hbar^2\gamma_N^2\Tr[\hat{I}_B^2]}\right]^{1/2}= \\
&=\frac{1}{4\sqrt{5}\hbar\gamma_N} \mid E_{Qz}(3\cos^2\theta-1)+\sqrt{3} E_{Q\perp}\sin^2\theta\cos2(\zeta-\alpha) \mid.
\end{split}}
\label{eq:B_Lq}
\end{equation}

The secular part of the spin-spin Hamiltonian reads \cite{clough1966nuclear}:
\begin{equation} \medmath{
\begin{split}
&\hat{H}_{ss}=\frac{1}{2} \hbar^2\gamma_{As}^2 \sum_{j>k} r_{jk}^{-3} \left(1-3\cos^2\theta_{jk} \right) \left( 3\hat{I}_{jz} \hat{I}_{kz}-\vec{\hat{I}}_j \vec{\hat{I}}_k \right)+\\
&+\frac{1}{2}\hbar^2\gamma_{As}\sum_{j>k'}\gamma_{k'Ga}\left[\hat{A}_{jk'}+(\hat{B}_{jk'}+r_{jk'}^{-3})(1-3\cos^2\theta_{jk'})\right]\hat{I}_{jz}\hat{I}_{k'z}.
\end{split}}
\label{eq:hamiltonian}
\end{equation}
Here indices $j$ and $k$ numerate the As atoms, and $k^\prime$ numerates the Ga atoms. Scalar and pseudodipole interactions are short-range, therefore, the constants $\hat{A}_{jk}$ and $\hat{B}_{jk}$ are not zero for the four Ga nuclei nearest to the $j$-th As nucleus only.
The spin-spin contributions to the local field for a \textsuperscript{75}As nucleus in the GaAs layer, neglecting the effect of interfaces, have cubic symmetry. As all the interactions are bi-linear in the spin operators, $B_{Lss}^2$, calculated along Eq.~\eqref{eq:local} as a function of polar angle $\theta$ and azimuthal angle $\alpha$, comprises cubic invariants of zeroth and fourth order:
\begin{equation}
B_{Lss}^2(\theta, \alpha)=C+D\left[\sin^4\theta(\cos^4\alpha+\sin^4\alpha)+\cos^4\theta\right],
\label{eq:B_Lss}
\end{equation}
where the directions of the cubic axis correspond to $\theta=0$ or $\theta=\pi$ and {$\theta=\pi/2$; $\alpha=n\pi/2$}. Numerical summation over the zinc-blend lattice of GaAs yields the constants $C$ and $D$ expressed via the parameters of the spin-spin Hamiltonian given by Eq.~\eqref{eq:hamiltonian}. Details of the calculation of the spin-spin and quadrupole contribution to the local field are given in the Appendix B.

\section{Sample and experimental technique}
\label{sec:experiment}
\begin{figure}
\center{\includegraphics[width=1\linewidth]{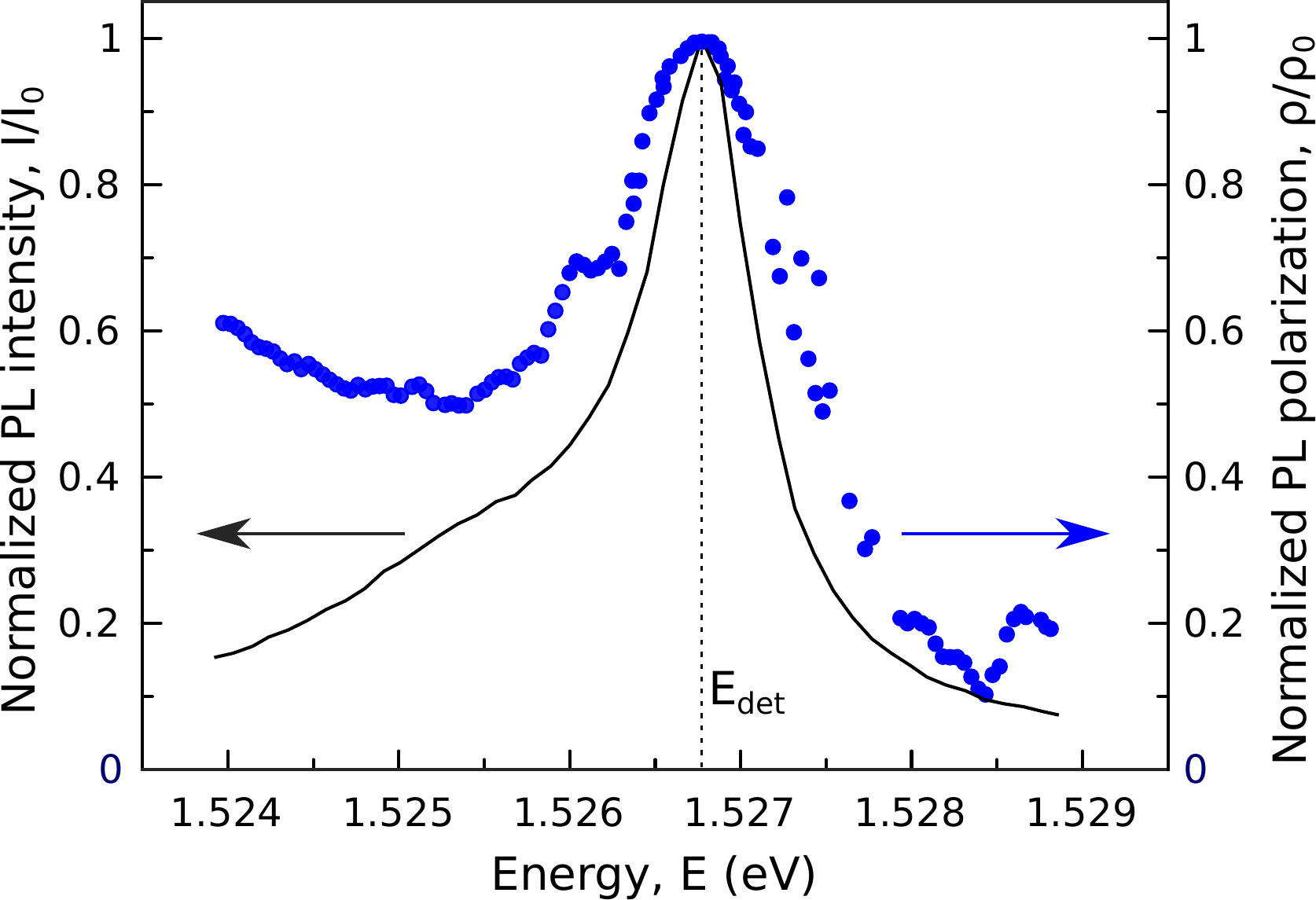}}
\caption{Photoluminescence spectra of 19.7 nm QW measured at $T=5.5$ K: intensity (black line) and circular polarization (blue circles). $E_{det}=1.5267$ eV indicates the chosen PL detection energy.}
\label{spectrum}
\end{figure}

The studied sample was grown by molecular beam epitaxy on a Te-doped GaAs substrate and consists of 13 nominally undoped GaAs/Al\textsubscript{0.35}Ga\textsubscript{0.65}As QWs with thicknesses varying from 2.8 nm to 39.3 nm separated by 30.9 nm thick barriers. The sample was placed in a He-flow cryostat and cooled down to $T=5.5$ K. For the optical excitation, when measuring the photoluminescence (PL) spectrum, circularly polarized light modulated in sign ($\sigma^+ / \sigma^-$) from a tunable diode laser was used with excitation energy $E_{exc}$ set to 1.5498 eV. The PL was collected in the reflection geometry, passed through a spectrometer and detected using an avalanche photodiode (APD). The PL intensity spectrum of the 19.7 nm QW and its circular polarization degree $\rho=\frac{I^+-I^-}{I^++I^-}$, where $I^+$ $(I^-)$ is the intensity of the right (left)-hand circularly polarized PL emission, is shown in Fig.~\ref{spectrum}. The maximum of the PL intensity detected at 1.5267 eV is attributed to the neutral exciton (X$^0$) emission \cite{mocek2017high}. It was chosen as a PL detection energy in all subsequently performed measurements. Accordingly, the excitation energy of the diode laser was tuned closer to resonance, $E_{exc}= 1.5276$ eV.
\begin{figure}
\center{\includegraphics[width=1\linewidth]{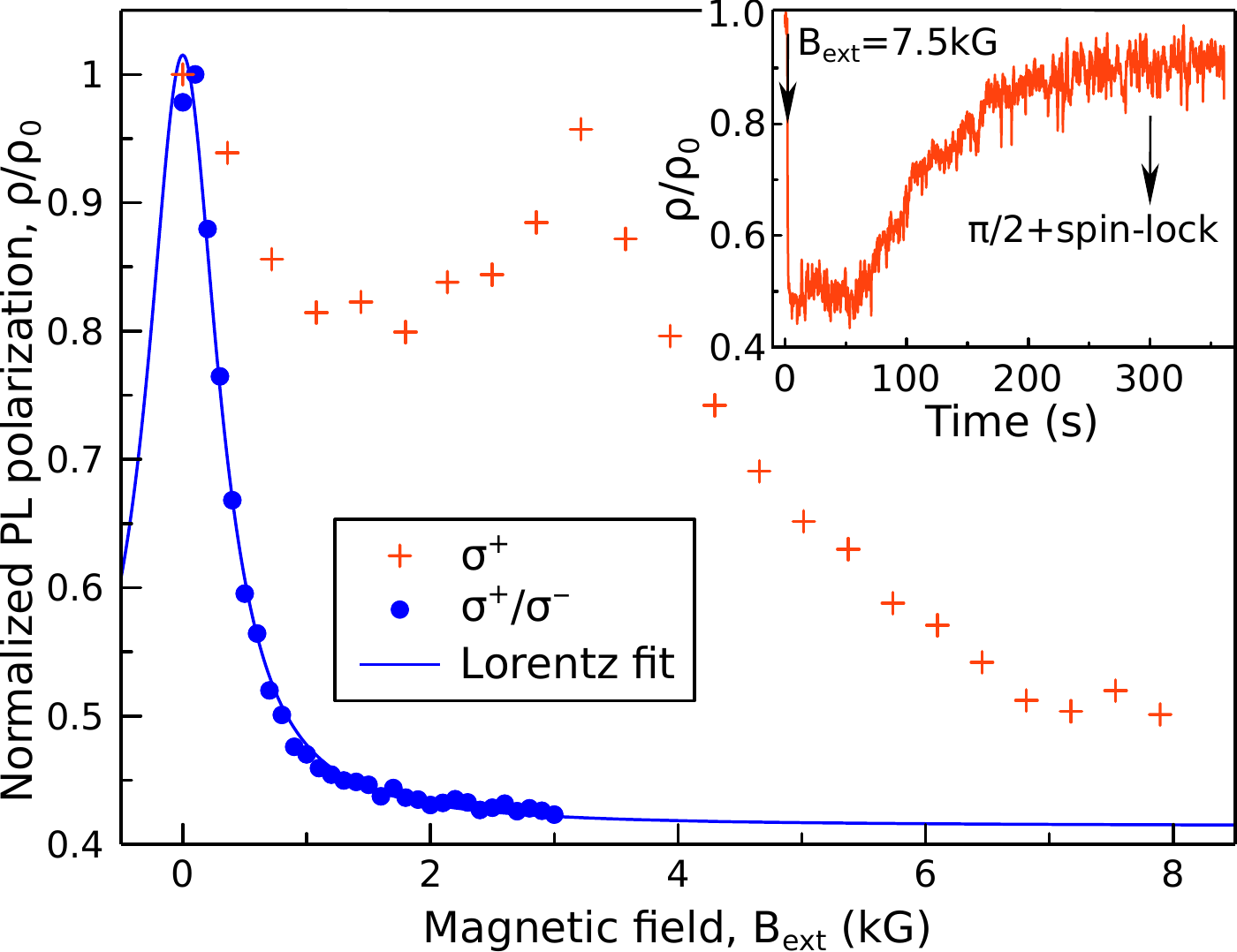}}
\caption{Depolarization of luminescence by oblique (66$^\circ$ from Faraday geometry) magnetic field with $\sigma^+/\sigma^-$ modulated at 50 kHz (blue circles) and constant $\sigma^+$ excitation (red crosses). The excitation and detection energies were equal to $E_{exc}=1.5276$ eV and $E_{det}=1.5267$ eV. The maximum of circular polarization degree at zero field amounted to $\rho_0\approx12$ \%. Inset: Time evolution of the nuclear spin polarization measured through the PL circular polarization degree at $B_{ext}=7.5$ kG. Arrows indicate the moment at which the external field and RF pulse sequence were applied.}
\label{Hanle}
\end{figure}
\begin{figure*}
\center{\includegraphics[width=0.97\linewidth]{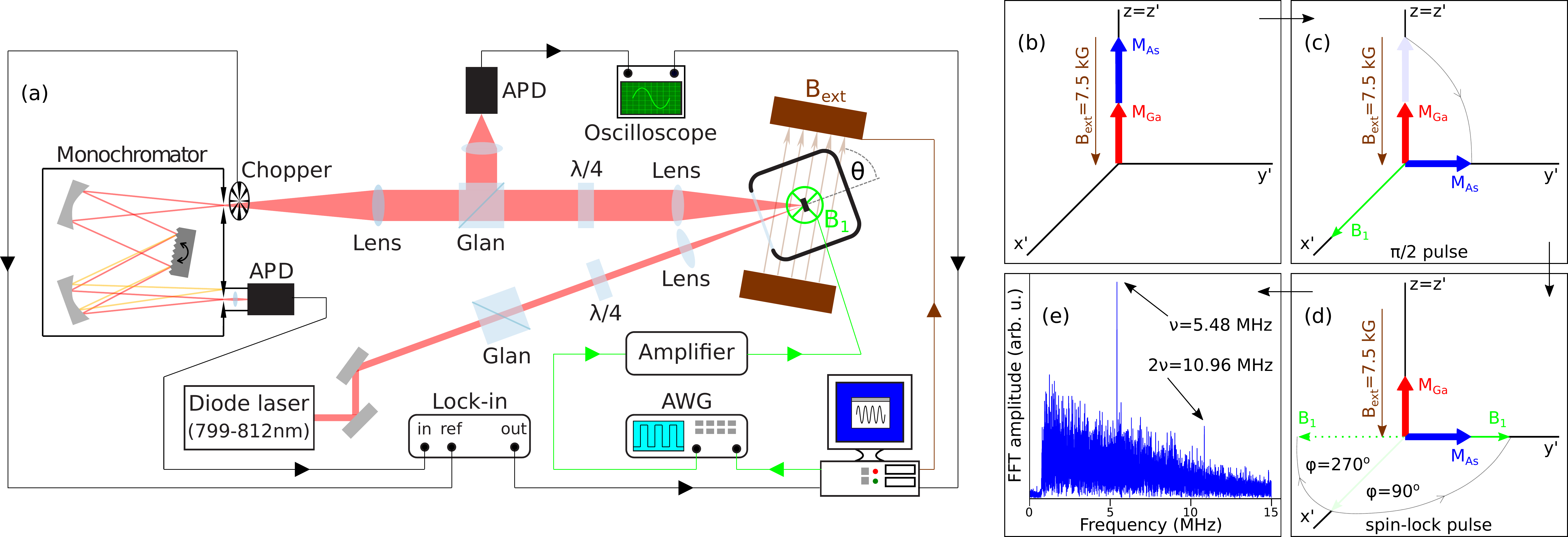}}
\caption{(a) Sketch of the experimental setup used for ADRF measurements. The experimental protocol presented on the right panel consisted of three stages: (b) During the first stage the sample was pumped with circularly polarized light in an external magnetic field, $B_{ext}=7.5$ kG. Helicity of light was chosen so that the resulting nuclear and external magnetic fields were anti-parallel. (c) The second stage begins with sending a $\pi/2$ pulse that tips the mean spin of \textsuperscript{75}As by $90^\circ$ in relation to the external magnetic field. (d) Immediately after the $\pi/2$ pulse, a spin-lock pulse was applied shifted in phase by $90^\circ$ ($270^\circ$) which makes the RF field parallel (antiparallel) to the tipped nuclear spin magnetization of \textsuperscript{75}As in the rotating frame. (e) In the third phase, after the end of the spin-lock pulse, the FID signal is measured and Fourier transformed from time to frequency domain.}
\label{setup}
\end{figure*}
The depolarization of the PL by the oblique (66$^\circ$ deviation from the structure axis) magnetic field $B_{ext}$ is shown in Fig.~\ref{Hanle}. When the sample was excited by the modulated ($\sigma^+/\sigma^-$ at 50 kHz) circularly polarized light, the transfer of spin polarization via the hyperfine interaction from optically oriented electrons to nuclei is hindered and depolarization of the electron spins in the external magnetic field (Hanle effect) is well represented by a Lorentzian function \cite{meier2012optical,dyakonov2008spin}. Otherwise, for excitation with fixed in sign circularly polarized light, the nuclei become polarized through hyperfine-induced flip-flop transitions of the electron and nuclear spins giving rise to the nuclear Overhauser field $B_N$. The Hanle curve is affected by the presence of the Overhauser field, manifested, depending on the helicity of the laser beam, by the appearance of an additional maximum where the two fields are anti-parallel ($B_{ext}-B_N$) or faster depolarization of the PL when the two fields are parallel ($B_{ext}+B_N$) \cite{meier2012optical,dyakonov2008spin}. In order to observe the build-up of the nuclear spin polarization at a relatively high external magnetic field, $B_{ext}=7.5$ kG, the helicity of the pump was chosen in such a way as to polarize nuclear spins along the external field (positive spin temperature of the Zeeman reservoir). Under this arrangement, the Overhauser field acting upon the electron spins from spin-polarized nuclei is anti-parallel to the external field. As it was shown for the same structure earlier \cite{kotur2018single}, in this case the nuclear spins are polarized efficiently until the compensation of the external magnetic field by the Overhauser field (see the inset in Fig~\ref{Hanle}).

The scheme of the experimental setup used in the ADRF measurements is shown in Fig.~\ref{setup}. The RF field at the sample was provided by the coils mounted on a cold finger of He-flow cryostat. The RF current passed through the coils was provided by the programmable arbitrary wavefunction generator (AWG) controlled by the computer. In this manner, it was possible to create a complex sequence of pulses with different amplitudes, phases or lengths. The experimental procedure consisted of three stages. In the first stage, nuclear spins were dynamically polarized via hyperfine interaction with electrons excited in the quantum well with circularly polarized pump light. In order to increase the net polarization of the selected isotope (\textsuperscript{75}As), spin polarization of the two Ga isotopes could be wiped out by sweeping the radiofrequency field across their resonance frequencies. This way, at the external field of 7.5 kG the polarization of the \textsuperscript{75}As spins theoretically could reach 28\%. The time duration of the first stage was determined by the dynamics of nuclear spin polarization at a given angle $\theta$ of the external field ($B_{ext}=7.5$ kG) and whether the erase RF sequence for the two Ga isotopes is applied. For $\theta=55^\circ$ and $\theta=66^\circ$, the sample was pumped for 300 s, sufficient for the Overhauser field to reach the saturation value of the external field. However, the pump period increased to 1000 s if the polarization of the \textsuperscript{69}Ga and \textsuperscript{71}Ga was suppressed by sending RF pulses at their resonant frequencies of 7.64 and 9.70 MHz, respectively. When $\theta=80^\circ$, saturation of the Overhauser field was reached after 1000 s, if the spin polarization of the Ga isotopes was not erased. The state of the Overhauser field was monitored through the PL component detected by the APD after passing through monochromator.

A second stage begins once the nuclear field reaches saturation, i.e. $B_N=-B_{ext}$, when the trigger signal is sent to the AWG and a sequence of radiofrequency pulses is applied (inset in Fig.~\ref{Hanle}). The $\frac{\pi}{2}$ pulse with the RF amplitude of 21 G turned the \textsuperscript{75}As mean spin perpendicular to the static 7.5 kG strong field. Then the locking RF field:
\begin{equation}
B_1(t)=2b\cos(\gamma_{As}B_{ext}t+\varphi)
\label{eq:B_1}
\end{equation}
was switched on, with the phase $\varphi$ shifted by 90$^\circ$ or 270$^\circ$ with respect to that of the tipping  pulse. In the coordinate frame rotating with the Larmor frequency of \textsuperscript{75}As ($\omega_L^{As}=\gamma_{As}B_{ext}$), one of the circular components of the locking field was static and directed parallel (positive spin temperature) or anti-parallel (negative spin temperature) to the tipped \textsuperscript{75}As mean spin $\langle I_{As} \rangle$. The initial amplitude of the locking field was 21 G, i.e. much stronger than the local nuclear field of 1.5-5 G (see below). We checked that there was no noticeable decay of the rotating \textsuperscript{75}As spin polarization during 3 seconds if the  locking field was kept on, while in the absence of the locking field the polarization decayed within $\approx200$ $\mu$s. This fact indicated that the \textsuperscript{75}As spin subsystem reached a thermodynamic equilibrium in the rotating frame, characterized by a spin temperature of a few  $\mu$K.  Then the locking field amplitude could be gradually changed to a varied final amplitude with the speed of $1\times10^4$ G/s, providing adiabatic de- or remagnetization in the rotating frame.

\begin{figure}
\center{\includegraphics[width=0.65\linewidth]{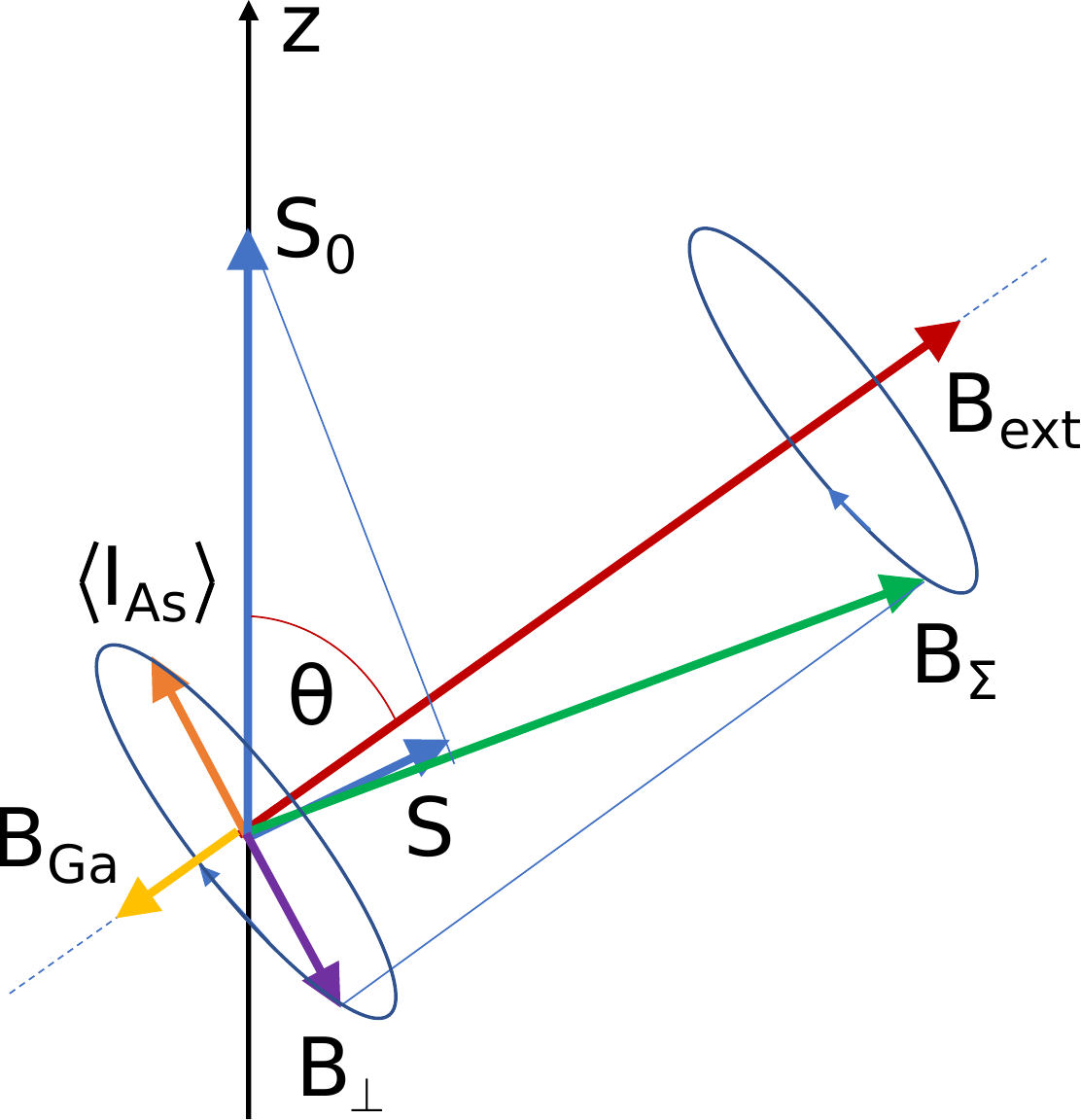}}
\caption{Graphical interpretation of the Hanle effect in a rotating nuclear field. As the Larmor period of the electron spin in the total fields $B_{\Sigma}$ is typically shorter than the electron spin lifetime, the electron mean spin $S$ is nearly parallel to $B_{\Sigma}$.  $S_Z$ oscillates while the field $B_{\perp}$, created by the tipped mean spin of \textsuperscript{75}As, rotates with the \textsuperscript{75}As Larmor frequency.}
\label{HanleRotate}
\end{figure}

In the third stage, the locking field was switched off and the free induction decay (FID) signal was recorded by measuring the PL circular polarization as a function of time with an assembly of a quaterwave plate and a Glan prism followed by a fast photodiode. The signal was Fourier-transformed and the amplitudes of 1\textsuperscript{st} and 2\textsuperscript{nd} harmonics of the \textsuperscript{75}As Larmor frequency were determined. The oscillating PL polarization used for FID detection results from the dynamic Hanle effect induced by the superposition of the static effective field $B_\Sigma$ (external field + Overhauser field from Ga isotopes, in case they were not wiped out during the first stage) and the rotating Overhauser field  produced by the tipped mean spin of \textsuperscript{75}As (see Fig.~\ref{HanleRotate}).

 As shown in Appendix A, the amplitudes of the 1\textsuperscript{st} and 2\textsuperscript{nd} harmonics of the oscillating PL polarization are equal to:
\begin{equation}
A_1=\frac{2T_2^*B_{\parallel}B_{\perp}(0)}{B_{\Sigma}^2}\sin\theta\cos\theta
\label{eq:A1}
\end{equation}
and
\begin{equation}
A_2=\frac{T_2^*B_{\perp}^2(0)}{4B_{\Sigma}^2}\sin^2\theta,
\label{eq:A2}
\end{equation}
respectively. Here $B_\Sigma^2=B_\parallel^2+B_\perp^2(0)$ is the rotating Overhauser field just after switching off the locking field, $T_2^*$ is the decay time of the rotating nuclear polarization, and $\theta$ is the angle of the external field to the structure axis. The ratio of amplitudes of the second and first harmonics equals to:
\begin{equation}
\frac{A_2}{A_1}=\frac{B_\perp(0)}{8B_\parallel}\tan\theta.
\label{eq:AmplitudeRatio}
\end{equation}
We used Eq.~\eqref{eq:AmplitudeRatio} to determine the initial value of the rotating Overhauser field of \textsuperscript{75}As in a strong, when compared to $B_L$, locking field, in order to obtain the absolute value of the spin temperature after ADRF. Since the 2\textsuperscript{nd} harmonic frequency falls slightly outside the 10 MHz bandwidth of our photodiode, it could not be reliably measured in case of weak $B_\perp$. For this reason, in the majority of experiments on ADRF only relative values of $B_\perp(0)$ were determined using Eq.~\eqref{eq:A1}.

\section{Results}
\label{sec:results}

\begin{figure}
\center{\includegraphics[width=1\linewidth]{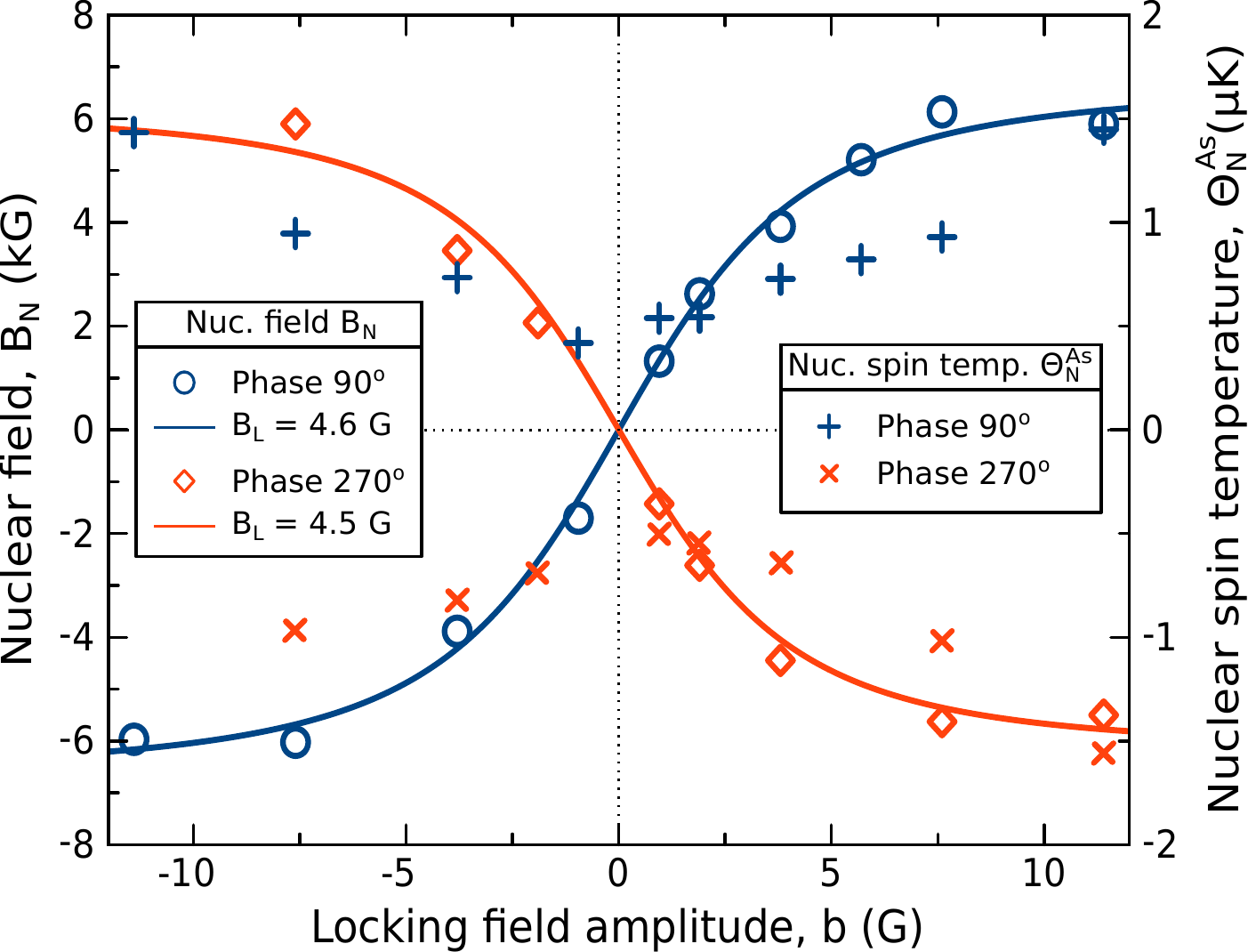}}
\caption{ADRF curves measured for $90^\circ$ (blue circles) and $270^\circ$ (red diamonds) phase shifts of the spin-lock from the $\frac{\pi}{2}$ pulse when the spin polarization of the two Ga isotopes was erased. The oblique external magnetic field $B_{ext}=7.5$ kG was directed at an angle of $\theta=66^\circ$. The solid lines are fits to Eq.~\eqref{eq:AmplitudeRatio} that determine
the values of $B_L$ and $\Theta_N^{As}$.}
\label{ADRM_As}
\end{figure}

The ADRF curves measured in the external magnetic field $B_{ext}=7.5$ kG directed at an angle of $\theta=66^\circ$ from Faraday geometry with spin-lock pulses shifted by $90^\circ$ or $270^\circ$ from the $\frac{\pi}{2}$ pulse are presented in Fig.~\ref{ADRM_As}. When the locking field differs in phase by $90^\circ$ ($270^\circ$) from the $\frac{\pi}{2}$ pulse, the adiabatic demagnetization process starts from a positive (negative) initial spin temperature. In this measurement, only \textsuperscript{75}As spins were pumped and the polarization of the two Ga isotopes was erased with RF pulses at their resonance frequencies. Fitting these curves with Eq.~\eqref{eq:B_perp} enables one to determine the values of the local fields for two different phases. Furthermore, knowing the value of the local field $B_L$ it is possible, using Eq.~\eqref{eq:temperature}, to obtain the value of nuclear spin temperature after adiabatic demagnetization to zero field ($b=0$):
\begin{equation}
\Theta_N^{As}(b=0)=\Theta_N^{As}(b_0)\frac{B_L}{\sqrt{b_0^2+B_L^2}},
\end{equation}
where $\Theta_N^{As}(b_0)=\hbar\gamma_{As}b_0I(I+1)/3\langle I \rangle$. $\langle I \rangle$ is determined from the nuclear field, created by rotating the As spins $\langle I \rangle=\hbar\gamma_eB_\perp(0)/A_{As}$, where $\gamma_e=3.52\times10^6$ radG\textsuperscript{-1}s\textsuperscript{-1} is the electron gyromagnetic ratio and $A_{As}=43.5$ $\mu$eV is the hyperfine coupling constant for \textsuperscript{75}As \cite{chekhovich2017measurement}. The calculated values of the nuclear spin temperature $\Theta_N^{As}(b=0)$ are given in Table \ref{tab:SpinTemperature}.

Similar measurements were performed for $\theta=80^\circ$ and $\theta=66^\circ$ without RF erase pulses for the Ga isotopes. In this context, all three isotopes contribute to the overall Overhauser field and the calculated values for the nuclear spin temperature after adiabatic demagnetization are made under the assumption that 52\% of the nuclear field value stems from spin polarized As nuclei and the other 48\% from the two gallium isotopes. This ratio is taken from the share of the three isotopes in the maximum possible nuclear field of 53 kG for GaAs under 100\% polarization \cite{meier2012optical}. For comparison, the values of nuclear spin temperature obtained this way are also added to Table \ref{tab:SpinTemperature}.

\begin{table}
\centering
\caption{Experimentally determined values of the nuclear spin temperature after adiabatic demagnetization to zero field.}
\begin{tabular}{c|c|c} \hline
$\theta$ ($^\circ$)& $\varphi$ ($^\circ$) &$\Theta_N^{As}(b=0)$ ($\mu$K) \\ \hline \hline
80 & 90 & 1.2 \\ \hline \hline
\multirow{4}{*}{66} & 90 & 1.6 \\ \cline{2-3}
& 270 & -2.1 \\ \hhline{~==}& 90 & 0.54\footnote{\label{ft}Spin polarization of the two Ga isotopes was erased.} \\ \cline{2-3}
& 270 & -0.57\footref{ft} \\ \hline
\end{tabular}
\label{tab:SpinTemperature}
\end{table}

The lowest spin temperatures of 2 $\mu$K \cite{vladimirova2018spin} and 5 $\mu$K \cite{kalevich1982onset} reported to date for semiconductors and semiconductor structures were measured in bulk GaAs by adiabatic demagnetization in the laboratory frame or at the set value of magnetic field $B_\perp=0.1$ G, respectively. Consequently, we have pushed the boundary of the accessible spin temperature range nearly four times down. However, these values represent the spin temperatures of all three isotopes, unlike in our case where we measure the spin temperature of \textsuperscript{75}As. Due to the absence of the strain induced quadrupole splitting of the nuclear spin states in bulk GaAs, the local field had the characteristic value of $B_L=B_{ss}=1.5$ G given by dipole-dipole interaction between the nuclei \cite{paget1977low}. On the other hand, the quadrupole splitting caused by strain leads to an increase in the local field, $B_L^2=B_{ss}^2+B_Q^2$, and consequently to a higher value of nuclear spin temperature, if the energy values of the quadrupole and spin-spin interactions are comparable \cite{vladimirova2018spin}, or to a breakdown of the nuclear spin temperature concept for strong quadrupole-induced local fields \cite{maletinsky2009breakdown}. In our QW sample, the spin-spin and quadrupole contributions to the local field are similar to each other so that the two reservoirs are effectively coupled and the nuclear spin system is characterized by a unique spin temperature. In spite of considerable quadrupole effects, this coupling has allowed us to enter the sub-microKelvin spin temperature range in a QW structure.

\begin{figure}
\center{\includegraphics[width=1\linewidth]{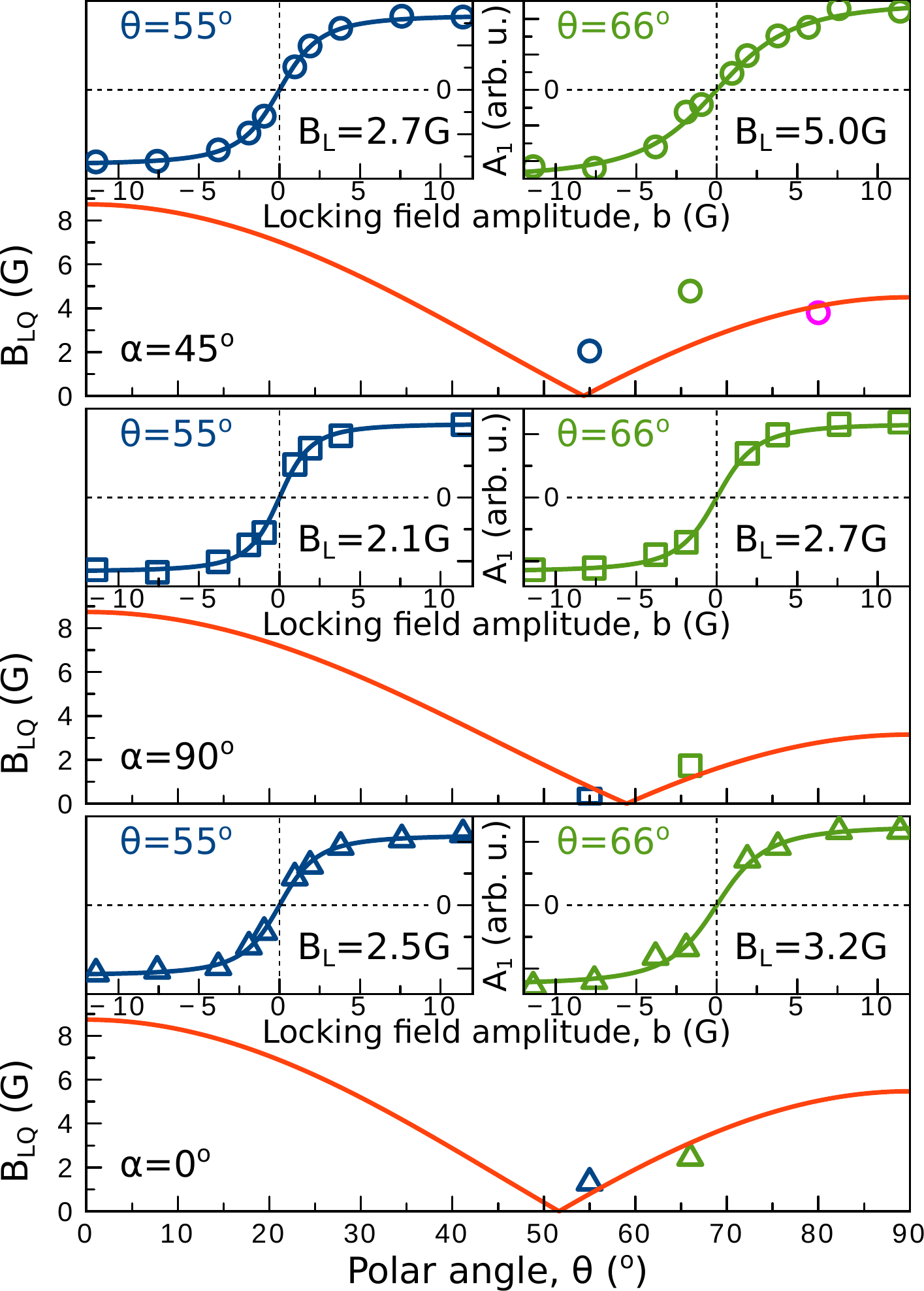}}
\caption{Dependence of the quadrupole part of the local field $B_{LQ}$ on the angle between the external magnetic field and the sample surface $\theta$ for three azimuths $\alpha$ ($45^\circ$, $90^\circ$ and $0^\circ$). The solid red lines are fits to Eq.~\eqref{eq:B_Lq}. Values for the $B_{LQ}$ were obtained by fitting the ADRF curves, measured for different polar and azimuthal angles and shown in the left and right insets, with Eq.~\eqref{eq:A2} (solid blue and green lines). Since for $\theta=80^\circ$ we only performed one measurement when the azimuthal angle $\alpha$ was equal to $45^\circ$, the obtained value for $B_{LQ}$ was added to the graph (pink circle) without showing the corresponding ADRF curve.}
\label{24and35degree}
\end{figure}

The dependence of the local magnetic field on the external magnetic field direction in and out of the sample plane was studied by measuring ADRF for $66^\circ$ and $55^\circ$ polar and $0^\circ$, $45^\circ$ and $90^\circ$ azimuthal angles. Considering the fact that the amplitude of the second harmonic $A_2$ for $\theta=55^\circ$ was too small to be determined from the measured FFT spectra, the dependence of the amplitude of the first harmonic $A_1$ on the magnitude of the RF field are presented in Fig.~\ref{24and35degree}. Using Eq.~\eqref{eq:A1} with $B_\Sigma^2=B_\parallel^2+B_\perp^2$, where $B_\perp$ is represented by Eq.~\eqref{eq:B_perp}, to fit the measured ADRF curves we get the values of the local field $B_L$. The measured local fields consist of two parts associated with the spin-spin and quadrupole interactions, $B_L=\sqrt{B_{Lss}^2+B_{LQ}^2}$. Using Eq.~\eqref{eq:B_Lss} to calculate the values for the spin-spin contribution to the local field $B_{Lss}$ it is possible to extract the values for the quadrupolar part $B_{LQ}$ for various polar and azimuthal angles which are presented in Fig. \ref{24and35degree}.

In order to compare our experimental data in the framework of the theoretical model, the obtained values for $B_{LQ}$ were fitted with Eq.~\eqref{eq:B_Lq}. For the two unknown parameters $E_{Qz}$ and $E_{Q\perp}$ we used $E_{Qz}=2\times10^{-11}$ eV and $E_{Q\perp}=3\times10^{-12}$ eV, while the best agreement with the theory was achieved when the angle between the axis of the EFG tensor in plane and the [100] crystal axis was equal to $\zeta=134^\circ$. It can be clearly seen that overall, the experimental results follow the general trend predicted by the theory presented in Section \ref{sec:theory}, i.e. $B_{LQ}$ has the highest values for $\alpha=45^\circ$, in other words, when the external magnetic field is directed along the [110] axis. When changing the azimuthal angle by $45^\circ$ clockwise ($\alpha=0^\circ$) or counterclockwise ($\alpha=90^\circ$) the value of the quadrupole local field decreases and reaches a minimum at $\theta\approx55^\circ$.

\section{Conclusions}
We have studied the process of deep cooling of the nuclear spins by adiabatic de(re)-magnetization in the rotating frame in a GaAs/(Al,Ga)As quantum well. Within this approach the nuclear spins were polarized in an oblique external magnetic field and the adiabatic transformation was achieved by sending an RF sequence, at frequency set for the \textsuperscript{75}As isotope, consisting of $\frac{\pi}{2}$ and "spin-lock" pulses. The RF field amplitude in the "spin-lock" pulse was gradually decreased down to zero. Following the change of the \textsuperscript{75}As spin polarization with this slowly varying RF field we were able to confirm that the nuclear spin temperature concept is still valid for our sample regardless of the presence of strain-induced nuclear quadrupole splitting manifested in an increase of the local magnetic field.

We have also experimentally demonstrated that, for a semiconductor structure, in the case of adiabatic demagnetization in the rotating frame, the local magnetic field, characterizing the heat capacity of the nuclear spin system, is dependent on the orientation of the external magnetic field with respect to the crystal and structure axes. The local magnetic field at different polar and azimuth angles was measured and compared with the theory taking into account both dipole-dipole interactions on the zinc-blend crystal lattice and quadrupole splitting due to strain and electric field along the structure axis. Although certain deviation from the predicted angular dependence of the local field was found, the overall agreement between the experiment and theory is satisfactory. Finally, it turned out possible to cool the coupled nuclear spin-spin and quadrupole reservoirs, characterized by a unified spin temperature, by adiabatic demagnetization in the rotating frame down to the sub-microKelvin range. The lowest thus far, in semiconductors, spin temperatures of +0.54 and -0.57 $\mu$K are reached. We consider this an important step towards realization of nuclear magnetic ordering, expected to appear at nuclear spin temperatures below 0.1 $\mu$K.

In our experiments, the nuclear spin temperature 10 million times lower in absolute value than that of the crystal lattice was reached. It is worth to compare this reduction factor with previous works on nuclear spin cooling in the solid state. The lowest temperatures reached so far, in the nanoKelvin range, were obtained in metals by a “brute force” method with the reduction factor of a hundred thousand, which required two-stage pre-cooling of conduction electrons to sub-milliKelvin temperatures \cite{oja1997nuclear}. Cooling of the nuclear spins in dielectrics to fractions of microKelvin \cite{chapellier1970production}, in spite using initial microwave hyperpolarization of the nuclear spins via the solid state effect, still demonstrated a reduction coefficient not exceeding one million. Distinct from those previous works, requiring unique purpose-built setups, we have realized spin cooling to sub-microKelvin temperatures in a sample that was held in a standard helium flow cryostat at temperature of 5.5 K. This advantage of our approach, which is a result of high efficiency of optical spin orientation in semiconductors, makes cooling nuclear spins to ultra-low temperatures much more feasible than before, opening ways to their applications in e.g. quantum simulators \cite{georgescu2014quantum}.

\section*{Acknowledgements}
This work was supported by the Deutsche Forschungsgemeinschaft within the International Collaborative Research Center TRR 160 (project A6) and Russian Foundation for Basic Research (project 19-52-12043). K.V.K. and V.M.L. acknowledge support from the St. Petersburg State University (research grant No. 73031758).

\section*{Appendix A: Hanle effect in rotating nuclear field}
\label{sec:AppendixA}

The characteristic relaxation times of the electron spin are much shorter than those of the nuclei. Therefore, one can consider the Hanle effect in a static oblique effective field \cite{meier2012optical}:
\begin{equation}
\frac{S_z}{S_0}=\frac{T_{s\parallel}(B_\Sigma)}{\tau}\cos^2\theta+\frac{T_{s\perp}}{\tau}\frac{1}{1+B_\Sigma^2/B_H^2}\sin^2\theta,
\label{eq:SzS0}
\end{equation}
where $S_z$ is the electron mean spin projection on the direction of light propagation in the sample, $\tau$ is the electron lifetime and $\theta$ is the angle between $B_\Sigma$ and $z$-axis. The lifetimes of spin components along and perpendicular to the effective field $B_\Sigma$ are given by the expressions:
\begin{equation}
\begin{split}
T_{s\parallel}(B_\Sigma)&=T_s(\infty)-\frac{T_s(\infty)-T_s(0)}{1+B_\Sigma^2/B_{PRC}^2} \\
T_{s\perp}&=T_s(0),
\end{split}
\end{equation}
where $B_{PRC}$ is the half width at half maximum (HWHM) of the polarization recovery curve (PRC) \cite{smirnov2020spin}, and $B_H$ is the HWHM of the Hanle curve. Introducing $B_z$, the component of $B_\Sigma$ along $z$, one can rewrite Eq.~\eqref{eq:SzS0} as:
\begin{equation} \medmath{
\begin{split}
&\frac{S_z}{S_0}=\frac{B_z^2}{B_\Sigma^2}\frac{T_{s\parallel}(B_\Sigma)}{\tau}+\frac{B_\Sigma^2-B_z^2}{B_\Sigma^2}\frac{T_{s\perp}}{\tau}\frac{1}{1+B_\Sigma^2/B_H^2}= \\
&=\frac{T_{s\perp}}{\tau}\frac{1}{1+B_\Sigma^2/B_H^2}+\frac{B_z^2}{B_\Sigma^2}\left[\frac{T_{s\parallel}(B_\Sigma)}{\tau}-\frac{T_{s\perp}}{\tau}\frac{1}{1+B_\Sigma^2/B_H^2}\right].
\end{split}}
\label{eq:SzS0_2}
\end{equation}
The effective field $B_\Sigma$ has two components: the static $B_\parallel$ $(B_\parallel \parallel B_{ext})$ and the rotating $B_\perp$ $(B_\perp \perp B_{ext})$. $B_z$ can be expressed via these components as follows:
\begin{equation}
B_z(t)=B_\parallel \cos\theta + B_\perp(t) \sin\theta \cos\omega_Nt,
\end{equation}
where $\omega_N$ is the Larmor frequency of the tipped isotope in the external field, $\theta$ is the angle between the external field and the $z$-axis and:
\begin{equation}
B_\perp(t)=B_\perp(0)e^{-t/T_2^*}.
\end{equation}
From Eq.~\eqref{eq:SzS0_2} we now obtain:
\begin{widetext}
\begin{equation}
\begin{split}
&\frac{S_z}{S_0}=\frac{T_{s\perp}}{\tau}\frac{1}{1+B_\Sigma^2/B_H^2}+\frac{B_\parallel^2\cos^2\theta}{B_\Sigma^2}\left[\frac{T_{s\parallel}(B_\Sigma)}{\tau}-\frac{T_{s\perp}}{\tau}\frac{1}{1+B_\Sigma^2/B_H^2}\right]+ \\
&+\left[\frac{T_{s\parallel}(B_\Sigma)}{\tau}-\frac{T_{s\perp}}{\tau}\frac{1}{1+B_\Sigma^2/B_H^2}\right] \left[\frac{B_\perp^2(t)}{2B_\Sigma^2}\sin^2\theta+\frac{B_\parallel B_\perp(t)}{B_\Sigma^2}\sin(2\theta)\cos(\omega_Nt)+\frac{B_\perp^2(t)}{2B_\Sigma^2}\sin^2\theta\cos(2\omega_Nt)\right].
\end{split}
\label{eq:SzS0_3}
\end{equation}
\end{widetext}
Eq.~\eqref{eq:SzS0_3} can be further rewritten as:
\begin{equation}
\frac{S_z}{S_0}=\Upsilon+\Xi(\omega_Nt),
\end{equation}
where:
\begin{equation} \medmath{
\begin{split}
&\Upsilon=\frac{T_{s\perp}}{\tau}\frac{1}{1+B_\Sigma^2/B_H^2}+ \\
&+\frac{2B_\parallel^2\cos^2\theta + B_\perp^2(t)\sin^2\theta}{2B_\Sigma^2}\left[\frac{T_{s\parallel}(B_\Sigma)}{\tau}-\frac{T_{s\perp}}{\tau}\frac{1}{1+B_\Sigma^2/B_H^2}\right]
\end{split}}
\end{equation}
and:
\begin{equation} \medmath{
\begin{split}
&\Xi(\omega_Nt)=\left[\frac{T_{s\parallel}(B_\Sigma)}{\tau}-\frac{T_{s\perp}}{\tau}\frac{1}{1+B_\Sigma^2/B_H^2}\right]\times \\
&\times\left[\frac{B_\parallel B_\perp(t)}{B_\Sigma^2}\sin(2\theta)\cos(\omega_Nt)+\frac{B_\perp^2(t)}{2B_\Sigma^2}\sin^2\theta\cos(2\omega_Nt)\right]
\end{split}}
\label{eq:Xi}
\end{equation}
are the non-oscillating and oscillating parts of the $z$-projection of the electron spin polarization.
Before the $\frac{\pi}{2}$ pulse and after the decay of free induction, $B_\perp=0$ and:
\begin{equation}
\frac{S_z}{S_0}=\frac{T_{s\parallel}(B_\Sigma)}{\tau}\cos^2\theta+\frac{T_{s\perp}}{\tau}\frac{1}{1+B_\Sigma^2/B_H^2}\sin^2\theta.
\end{equation}
Before the $\frac{\pi}{2}$ pulse $B_\Sigma$ corresponds to polarized nuclei (either all isotopes or only As). After the decay of free induction, $B_\Sigma$ is the same minus the contribution of As. During FID, the oscillating part $\Xi(\omega_Nt)$ has contributions at single and double Larmor frequency of As. Taking the Fourier transform of Eq.~\eqref{eq:Xi}, we find that their amplitudes are equal to:
\begin{equation}
A_1=\frac{2T_2^*B_\parallel B_\perp(0)}{B_\Sigma^2}\sin\theta\cos\theta
\end{equation}
and:
\begin{equation}
A_2=\frac{T_2^*B_\perp^2(0)}{4B_\Sigma^2}\sin^2\theta.
\end{equation}

The ratio of amplitudes of the second and first harmonics equals:
\begin{equation}
\frac{A_2}{A_1}=\frac{B_\perp(0)}{8B_\parallel}\tan\theta.
\end{equation}
Here $B_\perp(0)$ is the initial value of the rotating As field, and $B_\parallel$ is the static effective field (the same as remains after FID). Therefore, the amplitude of the rotating nuclear field can be determined from experiment as:
\begin{equation}
B_\perp(0)=8B_\parallel \frac{A_2}{A_1}\cot\theta=8B_\parallel \frac{A_2}{A_1}\tan\left(\frac{\pi}{2}-\theta\right).
\end{equation}
\\
\section*{Appendix B: Calculation of the spin-spin local field}
\label{sec:AppendixB}
\begin{figure}
\center{\includegraphics[width=0.9\linewidth]{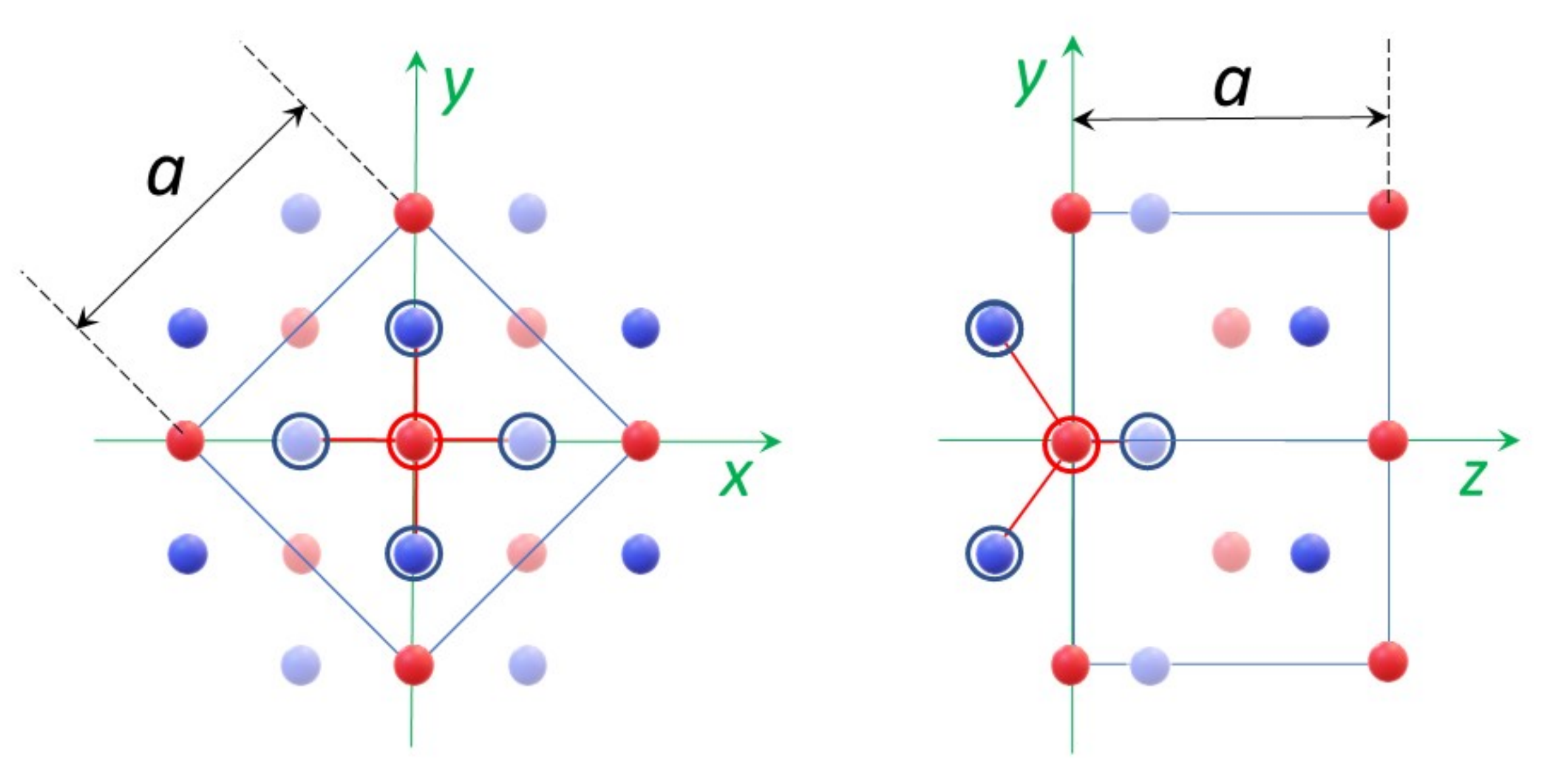}}
\caption{Fragment of the crystal lattice of GaAs and the coordinate frame used to calculate the spin-spin interactions for the selected As nucleus outlined by the red circle. Left: view along [001], right: view along [110]. Bright and pale red balls denote As nuclei with even and odd $k$, while bright and pale blue balls denote Ga nuclei with even and odd $k'$, correspondingly (see Eqs.~\eqref{eq:ind} and~\eqref{eq:ind_prime}). The cubic cell of the fcc As sublattice with the lattice constant $a$ is shown by blue lines. The four nearest Ga nuclei are outlined by blue circles.}
\label{lattice}
\end{figure}

The squared spin-spin local field:
\begin{equation}
B_L^2=\frac{\Tr[\hat{H}_{ss}^2]}{\hbar^2\gamma_{As}^2\Tr[\hat{I}_B^2]}
\end{equation}
comprises three contributions: from the As nuclei, $B_{As}^2$; from the four Ga nuclei next to the As nucleus, $B_{NGa}^2$; and from the rest of Ga nuclei, $B_{RGa}^2$. Since scalar and pseudo-dipole interactions are short-range, $B_{As}^2$ and $B_{RGa}^2$ are of purely dipole-dipole origin. In order to calculate these contributions, we introduce the Cartesian coordinate system with the axes $x\parallel[110]$, $y\parallel[1\bar{1}0]$, and $z\parallel[001]$ (see Fig. \ref{lattice}). The coordinates of the As and Ga nuclei in this frame are encoded by the indices $n$, $l$, $k$ for As:
\begin{equation} \medmath{
\begin{split}
x_{nk}^{As}&=a\left[\frac{\sqrt{2}}{2}n+\frac{\sqrt{2}}{4}\left(\frac{1-(-1)^k}{2}\right)    \right] \\
y_{lk}^{As}&=a\left[\frac{\sqrt{2}}{2}l+\frac{\sqrt{2}}{4}\left(\frac{1-(-1)^k}{2}\right)\right] \\
z_k^{As}&=a\frac{k}{2}
\end{split}}
\label{eq:ind}
\end{equation}
and $n’$, $l’$, $k’$ for Ga:
\begin{equation} \medmath{
\begin{split}
x_{n'k'}^{Ga}&=a\left[\frac{\sqrt{2}}{2}n'+\frac{\sqrt{2}}{4}\left(\frac{1-(-1)^{k'}}{2}\right)    \right] \\
y_{l'k'}^{Ga}&=a\left[\frac{\sqrt{2}}{2}l'+\frac{\sqrt{2}}{4}\left(\frac{1-(-1)^{k'}}{2}\right)\right] \\
z_{k'}^{Ga}&=a\left(\frac{k'}{2}-\frac{1}{4}\right).
\end{split}}
\label{eq:ind_prime}
\end{equation}
We start from calculation of the purely dipolar contributions, $B_{As}^2$ and $B_{RGa}^2$
\begin{equation} \medmath{
\begin{split}
&\hat{H}_{ss}=\frac{1}{2}\hbar^2\gamma_{As}^2\sum_{j>k}r_{jk}^{-3}(1-3\cos^2\theta_{jk})(3\hat{I}_{jz}\hat{I}_{kz}-\vec{\hat{I}}_{j}\vec{\hat{I}}_{k})+ \\
&+\frac{1}{2}\hbar^2\gamma_{As}\sum_{j>k}\gamma_{kGa}\left[\hat{A}_{jk}+(\hat{B}_{jk}+r_{jk}^{-3})(1-3\cos^2\theta_{jk})\right]\hat{I}_{jz}\hat{I}_{kz}.
\end{split}}
\end{equation}
Using Eqs.~\eqref{eq:local} and~\eqref{eq:hamiltonian} and applying the relation $\Tr \left[\hat{I}_{j\alpha}\hat{I}_{k\beta} \right]=\frac{I(I+1)(2I+1)}{3}\delta_{jk}\delta_{\alpha\beta}$ we find:
\begin{equation} \medmath{
\begin{split}
&B_{As}^2=\frac{1}{\hbar^2\gamma_{As}^2 \Tr\left\{\left[\sum{k}\hat{I}_{kB} \right]^2\right\}}\times \\
&\times\Tr\left\{\left[\frac{1}{2}\hbar^2\gamma_{As}^2\sum_{j>k}r_{jk}^{-3}\left(1-3\cos^2\theta_{jk}\right)\left(3\hat{I}_{jz}\hat{I}_{kz}-\vec{\hat{I}}_j\vec{\hat{I}}_k\right)\right]^2\right\}= \\
&=\frac{15}{2}\left(\frac{\hbar\gamma_{As}}{a^3}\right)^2 F_{As}(\theta, \alpha)\approx4.9\times10^{-3}[\textnormal{G}^2]F_{As}(\theta, \alpha)
\end{split}}
\end{equation}
and:
\begin{equation} \medmath{
\begin{split}
&B_{RGa}^2=\frac{1}{\hbar^2\gamma_{As}^2 \Tr\left\{\left[\sum{k}\hat{I}_{kB} \right]^2\right\}}\times \\
&\times\Tr\left\{\left[\frac{1}{2}\hbar^2\gamma_{As}^2\sum_{j>k}\gamma_{kGa}r_{jk}^{-3}\left(1-3\cos^2\theta_{jk}\right)\hat{I}_{jz}\hat{I}_{kz}\right]^2\right\}= \\
&=\frac{5}{4}\left[X_{69}\left(\frac{\hbar\gamma_{69Ga}}{a^3}\right)^2+X_{71}\left(\frac{\hbar\gamma_{71Ga}}{a^3}\right)^2\right] F_{Ga}(\theta, \alpha)\approx \\
&\approx1.99\times10^{-3}[\textnormal{G}^2]F_{Ga}(\theta, \alpha),
\end{split}}
\end{equation}
where $X_{69}$ and $X_{71}$ are the abundances of the corresponding Ga isotopes. The dimensionless functions $F_{As}(\theta,\alpha)$ and $F_{Ga}(\theta,\alpha)$ are defined as:
\begin{equation}
\begin{split}
&F_{As}(\theta,\alpha)=\frac{1}{2}\sum_{nlk}\frac{\left(1-3\cos^2\phi_{nlk}^{As}\right)^2}{\left(r_{nlk}/a\right)^6} \\
&F_{Ga}(\theta,\alpha)=\sum_{n'l'k'\notin NN}\frac{\left(1-3\cos^2\phi_{n'l'k'}^{Ga}\right)^2}{\left(r_{n'l'k'}/a\right)^6},
\end{split}
\end{equation}
where the indices $n,l,k$ and $n',l',k'$ are defined in Eqs.~\eqref{eq:ind} and~\eqref{eq:ind_prime}, $r_{nlk}$ or $r_{n'l'k'}$ stand for the distance to the corresponding nucleus from the selected As nucleus and $\phi_{nlk}^{As}$ or $\phi_{n'l'k'}^{Ga}$ denote the angle between the external field and the direction to this nucleus (see Fig.~\ref{lattice}).
Expressing the cosines of the angles $\phi_{nlk}$ and $\phi_{n'l'k'}$ via the coordinates of the corresponding nuclei and the direction cosines of the magnetic field:
\begin{equation} \medmath{
\begin{split}
&\cos\phi_{nlk}^{As}=\frac{x_{nk}^{As}\sin\theta\cos\alpha'+y_{lk}^{As}\sin\theta\sin\alpha'+z_k^{As}\cos\theta}{\sqrt{\left(x_{nk}^{As}\right)^2+\left(y_{lk}^{As}\right)^2+\left(z_{k}^{As}\right)^2}} \\
&\cos\phi_{n'l'k'}^{Ga}=\frac{x_{n'k'}^{Ga}\sin\theta\cos\alpha'+y_{l'k'}^{Ga}\sin\theta\sin\alpha'+z_{k'}^{Ga}\cos\theta}{\sqrt{\left(x_{nk}^{As}\right)^2+\left(y_{lk}^{As}\right)^2+\left(z_{k}^{As}\right)^2}},
\end{split}}
\end{equation}
where $\phi'=\phi+\frac{\pi}{4}$, we obtain:
\begin{widetext}
\begin{equation}
F_{As}(\theta,\alpha)=\frac{1}{2}\sum_{nlk}\frac{\left[x_{nk}^2/a^2+y_{lk}^2/a^2+z_{k}^2/a^2-3(\sin\theta\cos\alpha' \cdot x_{nk}/a + \sin\theta\sin\alpha' \cdot y_{lk}/a + \cos\theta \cdot z_{k}/a)^2 \right]^2}{\left(x_{nk}^2/a^2+y_{lk}^2/a^2+z_{k}^2/a^2\right)^5}
\label{eq:F_As}
\end{equation}
and
\begin{equation}
F_{Ga}(\theta,\alpha)=\sum_{n'l'k'}\frac{\left[x_{n'k'}^2/a^2+y_{l'k'}^2/a^2+z_{k'}^2/a^2-3(\sin\theta\cos\alpha' \cdot x_{n'k'}/a + \sin\theta\sin\alpha' \cdot y_{l'k'}/a + \cos\theta \cdot z_{k'}/a)^2 \right]^2}{\left(x_{n'k'}^2/a^2+y_{l'k'}^2/a^2+z_{k'}^2/a^2\right)^5}.
\label{eq:F_Ga}
\end{equation}
\end{widetext}
The functions $F_{As}(\theta,\alpha)$ and $F_{Ga}(\theta,\alpha)$ can be expressed through cubic invariants of zeroth and fourth order, $F(\theta,\alpha)=C+D\left[\sin^4\theta(\cos^4\alpha+\sin^4\alpha)+\cos^4\theta\right]$, where the coefficients $C$ and $D$ can be determined from the numerical values of $F_{As}(\theta,\alpha)$ and $F_{Ga}(\theta,\alpha)$ for two pairs of angles, e.g.:
\begin{equation}
\begin{gathered}
C=2F(\pi/2,\pi/4)-F(0,0) \\
D=2\left[F(0,0)-F(\pi/2,\pi/4)\right].
\end{gathered}
\end{equation}
This way, we find:
\begin{equation}
\begin{gathered}
C_{As}=127.8; \;C_{Ga}=32.0 \\
D_{As}=-58.9; \; D_{Ga}=18.3.
\end{gathered}
\end{equation}
Finally, the contribution of the nearest four Ga nuclei can be calculated analytically:
\begin{widetext}
\begin{equation} \medmath{
\begin{split}
B_{N\!NGa}^2&=\frac{1}{\hbar^2\gamma_{As}^2}\Tr\left\{\left[\frac{1}{2}\hbar^2\gamma_{As}\sum_{k'\in N\!N} \gamma_{kGa}\left[\tilde{A}+(\tilde{B}+r_{k'}^{-3})(1-3\cos^2\phi_{k'})  \right]\hat{I}_{k'z}\right]^2\right\}= \\
&=\frac{5}{4}\left[X_{69}\left(\frac{\hbar\gamma_{69Ga}}{a^3}\right)^2+X_{71}\left(\frac{\hbar\gamma_{71Ga}}{a^3} \right)^2 \right]\sum_{k'\in N\!N}\left[b_{sc}+(b_{pd}+b_{dd})(1-3\cos^2\phi_{k'}) \right]^2= \\
&=\frac{5}{4}\left[X_{69}\left(\frac{\hbar\gamma_{69Ga}}{a^3} \right)^2+X_{71}\left(\frac{\hbar\gamma_{71Ga}}{a^3}\right)^2 \right]\times 4b_{sc}^2+8(b_{pd}+b_{dd})^2-8(b_{pd}+b_{dd})^2\left[\sin^4\theta(\sin^4\varphi+\cos^4\varphi)+\cos^4\theta \right]\approx \\
& \approx 1.99\times10^{-3}[\textnormal{G}^2] 4b_{sc}^2+8(b_{pd}+b_{dd})^2-8(b_{pd}+b_{dd})^2\left[\sin^4\theta(\sin^4\varphi+\cos^4\varphi)+\cos^4\theta\right],
\end{split}}
\end{equation}
\end{widetext}
where $b_{sc}=\tilde{A}a^3$, $b_{pd}=\tilde{B}a^3$ and $b_{dd}=(4/\sqrt{3})^3$. According to Ref.~\cite{cueman1976pseudodipolar}, $b_{sc}\approx0.6b_{dd}$ and $b_{pd}\approx-0.55b_{dd}$.

\bibliography{Bibliography}

\end{document}